\documentclass[preprint,aps,amssymb,showpacs,superscriptaddress]{revtex4}
\usepackage{graphicx}

\begin{document}
\preprint{IMSc/2003/07/17}
%\preprint{CGPG--03/07-- }
%\preprint{gr--qc/0307083}

\title{Consistency Conditions for Fundamentally Discrete Theories}

\author{Martin Bojowald}
\email{bojowald@gravity.psu.edu}
\affiliation{Center for Gravitational Physics and Geometry,\\
The Pennsylvania State University,\\
104 Davey Lab, University Park, PA 16802, USA}
\author{Ghanashyam Date}
\email{shyam@imsc.res.in}
\affiliation{The Institute of Mathematical Sciences\\
CIT Campus, Chennai-600 113, INDIA.}

\newtheorem{Def}{Definition}
\newtheorem{Prop}{Proposition}

\newcommand{\Case}[2]{{\textstyle \frac{#1}{#2}}}
\newcommand{\lP}{\ell_{\mathrm P}}
\newcommand{\HE}{H^{(\mathrm E)}}
\newcommand{\hatHE}{\hat{H}^{(\mathrm E)}}

\newcommand{\md}{{\mathrm{d}}}
\newcommand{\tr}{\mathop{\mathrm{tr}}}
\newcommand{\sgn}{\mathop{\mathrm{sgn}}}

\begin{abstract}
  The dynamics of physical theories is usually described by
  differential equations. Difference equations then appear mainly as
  an approximation which can be used for a numerical analysis. As
  such, they have to fulfill certain conditions to ensure that the
  numerical solutions can reliably be used as approximations to
  solutions of the differential equation. There are, however, also
  systems where a difference equation is deemed to be fundamental,
  mainly in the context of quantum gravity. Since difference equations
  in general are harder to solve analytically than differential
  equations, it can be helpful to introduce an approximating
  differential equation as a continuum approximation.  In this paper
  implications of this change in view point are analyzed to derive the
  conditions that the difference equation should satisfy.  The
  difference equation in such a situation cannot be chosen freely but
  must be derived from a fundamental theory. Thus, the conditions for
  a discrete formulation can be translated into conditions for
  acceptable quantizations. In the main example, loop quantum
  cosmology, we show that the conditions are restrictive and serve as
  a selection criterion among possible quantization choices.
\end{abstract}

\pacs{0460P, 0460K, 9880Q}

\maketitle

%%%%%%%%%%%%%%%%%%%%%%%%%%%%%%%%%%%%%%%%%%%%%%%%%%%%%%%%%%%%%%%%%%%%%%%

\section{Introduction}

While differential equations play an outstanding role throughout
physics, it has long been expected that a quantization of
gravitational systems will lead to a discrete structure of
space-time. Then, the differential equations must be replaced by
difference equations at a fundamental level. The process of
discretization itself is, in fact, well-known since it is used to
study differential equations numerically. In this case the difference
equation is only an approximation to the differential equation which
is better suited to a numerical analysis. When the fundamental
equation is discrete, however, the point of view must change. Now,
analytic calculations are more complicated and less exact techniques
are available than would be for a differential equation. Thus, it
becomes convenient to consider, in certain regimes, a differential
equation as an approximation to the difference equation. This is also
necessary since the continuum formulation gives an excellent
description of the real world such that there must be a continuum
approximation to the discrete theory.

In both cases the relation between a difference and a differential
equation is analyzed. But there is an important difference: In
numerics \cite{Numerics} one has a large freedom in choosing the
difference equation which can be exploited to obtain a very good
approximation to the differential equation with helpful properties for
the numerical analysis. If the difference equation is fundamental,
e.g.\ if it follows from a basic quantum theory, then it cannot be
changed so freely. Only quantization choices and ambiguities can be
exploited which are often more limited and have to respect other
consistency conditions of the theory. In particular, while the order
of the difference equation should not be less than that of the
differential equation to allow the same freedom as in the continuum
formulation, there is no guarantee that it is not of considerably
higher order. Also in numerics higher order schemes are being used to
improve the numerical results, but one has to choose the higher order
discretization very carefully. Whenever one uses a higher order, there
are additional solutions, unwanted from the numerical point of view,
which have to remain suppressed throughout the evolution. When the
fundamental difference equation is of higher order than a continuum
differential equation, there will also be additional solutions which
correspond to a new freedom of the theory not seen in the continuum
picture. While those solutions can play a physical role in certain
regimes, their role must be limited in semiclassical regimes for the
continuum formulation to be valid. Now, one has to check if the
freedom allowed by different quantization choices can not only provide
a discrete equation which has the expected continuum approximation,
but also, in the case of higher order, does not lead to growing modes
which would soon dominate the continuum approximations.

The aim of this paper is to extract conditions for a fundamental
difference equation to have a good continuum picture in this
sense. Moreover, we will discuss how these conditions can be used as
selection criteria among different quantizations or ambiguities. It
turns out that they can be quite restrictive, thus having implications
for quantization issues which at first glance seem only indirectly
related to the fundamental law.

Our motivation, and the main example for a theory where in fact a
discrete fundamental evolution law has been derived, is loop quantum
cosmology. The issue of discretization itself as a consequence of
quantum theory or gravity has been studied before \cite{Disc}, but in
particular the issue of higher order has not always been
appreciated. In order to cover also other examples of a discrete
formulation, we will hold the discussion more general, but will use
the example of loop quantum cosmology as an illustration throughout
the paper. Moreover, we expect that loop quantum cosmology, or more
generally quantum geometry which it is a part of, will be the prime
application of the selection criteria. The reason is that quantum
geometry is the most advanced formulation which has led to a discrete
structure. Its kinematical setup is at a mathematically precise and
well-defined level which demonstrates the main quantization
ambiguities involved in the definition of the basic laws, mainly the
Hamiltonian constraint. In such a scenario the selection criteria will
be most useful since they can be tested explicitly and can restrict
the possible choices. While explicit calculations in the full theory
are still outside current reach, one can study the basic law
explicitly in homogeneous models. Most of them, those with non-zero
intrinsic curvature, require an additional input compared to the full
theory such that the contact is not so close and more ambiguities
arise \cite{HomSpin}. It is then helpful to use the criteria discussed
here in order to select a quantization.

In section II we specify the class of difference equations considered.
We briefly recall the quantization of homogeneous models and describe
the form of the ensuing fundamental difference equations.

In section III, we discuss the intuitive notion of continuum
approximation and its formulation in terms of pre-classical
solutions. For definiteness and clarity, in this section we discuss
fundamental equations which are ordinary difference equations. The
notion of pre-classicality is carefully defined without appealing to
any limiting procedure (Def.\ \ref{PreClass}). Its implications for
the difference equation are detailed. Subsequently a constructive
procedure is given to demonstrate the existence of pre-classical
solutions. This procedure naturally leads to a notion of local
stability which is defined here (Def.\ \ref{DefLocStab}). The conditions
of admissibility of a pre-classical solution together with local
stability are given in terms of the coefficients of the fundamental
difference equation. These are our main results summarized in Props.\
\ref{Cond} and \ref{Roots}.

In section IV, the case of a fundamental partial difference equation
is analyzed. The analysis closely follows the steps detailed in
section III. In some cases the fundamental equation can be interpreted
as a non-homogeneous ordinary difference equation and this can be done
in several ways. For each of these, the definition of local stability
from section III can be applied. The fundamental difference equation
can then be constrained by requiring that the local stability criteria
must be satisfied for any evolutionary interpretation. The example of
homogeneous loop quantum cosmology \cite{HomSpin} is described
briefly.

In the concluding section V, we summarize our results. 

An appendix is included which illustrates an example of an older,
unstable quantization of isotropic loop quantum cosmology.

%%%%%%%%%%%%%%%%%%%%%%%%%%%%%%%%%%%%%%%%%%%%%%%%%%%%%%%%%%%%%%%%%%%%%%%

\section{Difference equations}

In this section we fix our general notation of the class of difference
equations considered, and describe the more special way in which they
arise from loop quantum cosmology.

\subsection{General set-up}

We assume that there is a continuum formulation for the system under
consideration which has a number of continuous parameters $p^I$ (which
can be space or space-time coordinates, or in the case of cosmology
minisuperspace coordinates) and a linear, homogeneous differential
equation ${\cal D}s=0$ of order $\kappa$ for a (wave) function
$s(p^I)$. These basic variables have an underlying discrete structure
given by the lattice
\begin{equation}
 p^I(m_I)=\gamma\ell m_I
\end{equation}
where $m_I$ are integers and $\gamma$ is the dimensionless parameter
which controls the scale of the discreteness. The parameter $\ell$
will always be fixed and can absorb possible dimensions. At the
discrete level, the fundamental equation is a difference equation of
the form
\begin{equation} \label{Gen}
 \sum_{\{i_I=-k\}}^k A_{\{i_I\}}^{\{m_I\}}(\gamma)\;s_{\{m_I+i_I\}}=0
\end{equation}
of order $2k$ for a (wave) function $s_{\{m_I\}}$ living on the
lattice (the notation $\{\ldots\}$ in the sum indicates that all $i_I$
are summed over). The coefficients can in general depend on the
coordinates $m_I$ as well as the parameter $\gamma$.

Due to the relation between $p^I$ and $m_I$ we expect a relation
between the discrete and the continuous formulation when $\gamma$ is
small and the $m_I$ are large, which will be assumed from now on. If
$\gamma$ is small but non-zero, the differential equation can only be
an approximation to the difference equation; a strict relation could
be achieved in the limit $\gamma\to0$, $m_I\to\infty$ such that $p^I$
is unchanged, which we will call, as in loop quantum cosmology, the
pre-classical limit. Since a relation between the formulations will be
demonstrated by using expansions
\begin{equation}
 s_{\{m_I+i_I\}}=s(p^I(m_I))+\sum_I\gamma i_I\partial_Is(p^I(m_I))+\cdots
\end{equation}
one also has to assume properties of the wave function concerning its
oscillation on small scales. This will be discussed in more detail
later.

We will also assume that the coefficients of the difference equation
can be expanded in terms of $\gamma$,
\begin{equation} \label{Aexp}
 A_{\{i_I\}}^{\{m_I\}}(\gamma)=\sum_{j=-l}^{\infty} \gamma^j
 A^{(j)}_{\{i_I\}}(p^I(m_I))
\end{equation}
where poles at most of order $l$ appear. (Some, but not all of the
$A^{(-l)}_{\{i_I\}}(p^I)$ can be zero.) Note that we use the $p^I$ to
parameterize the coefficients in the expansion. This is necessary
because later we will use the expansion in the continuum limit where
some of the $\gamma$-dependence is absorbed in the $m_I$. Furthermore,
we assume the following conditions for the coefficients to be
satisfied at least for large $m_I$ which would be used in a
pre-classical limit or approximation (at small $m_I$, the conditions
can be violated by quantum effects):
\begin{eqnarray}
 \mbox{unitarity conditions } &&
 A_{\{i_I\}}^{\{m_I\}}=\left(A_{\{-i_I\}}^{\{m_I\}}\right)^*\\
 \mbox{parity conditions } && A_{\{i_I\}}^{\{m_I\}}=\pm
 A_{\{i_I\}}^{\{-m_I\}}\,.
\end{eqnarray}
The first condition ensures that the equation with a reversal of the
direction of ``time'' is solved by the complex conjugate of the
original wave function:
\[
 \sum_{\{i_I=-k\}}^k A_{\{i_I\}}^{\{m_I\}}s_{\{m_I-i_I\}}^*=
 \sum_{\{i_I=-k\}}^k A_{\{-i_I\}}^{\{m_I\}}s_{\{m_I+i_I\}}^*=
 \left(\sum_{\{i_I=-k\}}^k A_{\{i_I\}}^{\{m_I\}}s_{\{m_I+i_I\}}\right)^*=
 0
\]
while the second one guarantees that a solution remains a solution
when all the coordinates are reversed.

Depending on the situation and the physical meaning of the parameters
$p^I$, we will usually assume rotational symmetry of the equations,
i.e.\ permutation symmetry for the coefficients $m_I$.

\subsection{Loop quantum cosmology}
\label{s:LQC}

In diagonal homogeneous loop quantum cosmology \cite{HomCosmo} we have
three independent variables $p^I$, $I=1,\ldots,3$, which are the
components of the densitized triad. They determine the spatial metric
as $\md s^2 =a_I^2(\omega^I)^2$ where $\omega^I$ are fixed invariant
1-forms which specify the particular model, and
$a_I=\sqrt{|p^Jp^K/p^I|}$ if $\epsilon_{IJK}=1$. If we restrict to
isotropic models, there is a single $p$ and $a=\sqrt{|p|}$ is the
scale factor.  The discreteness is controlled by the Barbero--Immirzi
parameter $\gamma$ of quantum geometry, and $\ell=\frac{1}{2}\lP^2$
with the Planck length $\lP$ (or $\ell=\frac{1}{6}\lP^2$ in the
isotropic case). Note that the regular lattice discretization
$p^I(m_I)=\frac{1}{2}\gamma\lP^2 m_I$ is not an assumption but derived
from the theory via the spectra of densitized triad operators; it is a
general feature of quantum geometry. For diagonal homogeneous models
the lattice structure follows from the eigenvalues $p^I(m_I)$ of triad
operators $\hat{p}^I$ with eigenstates $|m_1,m_2,m_3\rangle$. A wave
function is supported on this lattice, $s_{m_1,m_2,m_3}$ in the triad
representation defined by $|s\rangle=\sum_{\{m_I\}}
s_{m_1,m_2,m_3}|m_1,m_2,m_3\rangle$ for an arbitrary state
$|s\rangle$. Since the states $|m_1,m_2,m_3\rangle$, defined as
eigenstates of the triad operators, have the freedom of an arbitrary
(possibly $m_I$-dependent) phase factor, the triad representation and
the fundamental equation for a wave function are fixed only after a
choice of the phase factors has been made.

Another essential part of quantum geometry is that the densitized
triad is conjugate to a connection, which in the diagonal homogeneous
case also has three components $c_I=\Gamma_I-\gamma K_I$ where
$\Gamma_I$ is the spin connection and $K_I$ the extrinsic
curvature. It is quantized in quantum geometry not directly, but via
its holonomies \cite{Bohr}. In the homogeneous case, these holonomies
are given essentially by $\sin(\frac{1}{2}c_I)$ and
$\cos(\frac{1}{2}c_I)$ acting as multiplication operators in the
connection representation. If we choose the phases for the triad
eigenvectors as
\begin{equation}\label{cm}
 \langle c_1,c_2,c_3|m_1,m_2,m_3\rangle = 2^{-\frac{3}{2}}
 \frac{\exp(i(m_1c_1+m_2c_2+m_3c_3)/2)}{\sin(c_1/2)\sin(c_2/2)\sin(c_3/2)}
\end{equation}
the holonomy operators act on a wave function $s_{\{m_I\}}$ via,
e.g.,
\begin{eqnarray}
  (\sin(\Case{1}{2}c_1)s)_{m_1,m_2,m_3} &=&
  \Case{1}{2}i(s_{m_1+1,m_2,m_3}- s_{m_1-1,m_2,m_3}) \label{sin} \\
  (\cos(\Case{1}{2}c_1)s)_{m_1,m_2,m_3} &=&
  \Case{1}{2}(s_{m_1+1,m_2,m_3}+ s_{m_1-1,m_2,m_3})\;. \label{cos}
\end{eqnarray}

The fundamental equation is given by the Hamiltonian constraint
equation for a wave function. In the full theory, a part of it is
constructed with a Wilson loop evaluated in the connection. It can
straightforwardly be specialized to a homogeneous model provided that
there is no intrinsic curvature (i.e., the Bianchi I model). To
respect the symmetry we are only allowed to use integral curves of
vector fields generating the symmetry (dual to the invariant 1-forms
used above) in forming the Wilson loop; this can be done by forming a
square from four such curves which close because the vector fields
commute. The Hamiltonian constraint equation for the Bianchi I model
is then essentially fixed taking the form (\ref{DiffEq}) with non-zero
coefficients
\begin{equation} \label{HomCosmoCoeff}
 A_{2\epsilon_1,2\epsilon_2,0}^{m_1,m_2,m_3}(\gamma)=
 \epsilon_1\epsilon_2 \, \gamma^{-3}\kappa^{-1}\lP^{-2} 
 (V_{m_1-2\epsilon_1,m_2-2\epsilon_2,m_3+1}-
 V_{m_1-2\epsilon_1,m_2-2\epsilon_2,m_3-1})
\end{equation}
for $\epsilon_I=\pm 1$, and similarly for other coefficients by
permutation symmetry. We used the volume eigenvalues
\begin{equation}
 V_{m_1,m_2,m_3}=(\Case{1}{2}\gamma\lP^2)^{\frac{3}{2}}\sqrt{|m_1m_2m_3|}
\end{equation}
to simplify the notation, $\lP^2 = \kappa \hbar $, and $\kappa=8\pi G$ is the 
gravitational constant. If there is matter or a cosmological constant, there
will also be a coefficient (it can be operator valued, acting on the matter
dependence of the wave function)
\begin{equation}
 A_{0,0,0}^{m_1,m_2,m_3}(\gamma)=\hat{H}_{\rm matter}(m_1,m_2,m_3)\,.
\end{equation}
In the notation of (\ref{Aexp}) we have a pole of order $l=2$ and
non-vanishing coefficients which can be read off from
\begin{eqnarray*}
 V_{m_1-2\epsilon_1,m_2-2\epsilon_2,m_3+1} & - &
 V_{m_1-2\epsilon_1,m_2-2\epsilon_2,m_3-1} \\
 &=&
 \sqrt{|p^1-\epsilon_1\gamma\lP^2| |p^2-\epsilon_2\gamma\lP^2|}
 \left(\sqrt{p^3+\Case{1}{2}\gamma\lP^2}-
 \sqrt{p^3-\Case{1}{2}\gamma\lP^2}\right)\\
 &=& \Case{1}{2}\gamma\lP^2\sqrt{|p^1p^2/p^3|}
 \left(1-\Case{1}{2}\gamma\lP^2(\epsilon_1/p^1+\epsilon_2/p^2)
 +O(\gamma^2)\right)\,.
\end{eqnarray*}
The order of the partial difference equation is $2k=4$.

The fact that we have a difference equation can be traced back to the
Wilson loop operator which can be decomposed into a product of the
operators (\ref{sin}), (\ref{cos}) giving us the basic difference and
mean. It also demonstrates the point that a fundamental difference
equation can easily be of higher order: the basic difference
(\ref{sin}) is a second order difference operator, rather than first
order (corresponding to a leap-frog scheme in numerics). It is a
basic result of the theory and cannot be changed to simplify the
equations.

In other Bianchi models with non-zero intrinsic curvature the
situation is not so clear-cut. Their symmetries are generated by
vector fields which do not commute, thus not forming closed loops in
the way described above. We have to correct for this fact by making
another choice which {\em cannot\/} be motivated from the full
theory. There needs to be an additional step because non-zero
intrinsic curvature in a model is realized by a non-zero spin
connection. While this is an invariant statement in the model, it does
not make sense in the full theory where the spin connection can
be made to vanish locally. Therefore, it does not pose a problem in
the full theory, but we have to face it in any model where it does not
vanish \cite{HomSpin}. Guidance by the full theory, which we could use
for the Bianchi I model, is lost which results in more quantization
choices and ambiguities.

The purpose of the conditions studied in this paper is two-fold. We
have to check if they are fulfilled in the Bianchi I model which then,
due to its close relation to the full theory, can be regarded as a
consistency check for the full quantization. For the other models we
can use the conditions to select among the possibilities for a
quantization (if they are possible to realize at all).

\subsection{Pre-classical approximation vs.\ pre-classical limit}

It is obvious from the basic definitions that the smaller the
parameter $\gamma$, the closer we are to a continuum formulation. This
also requires to have large values of the discrete variables $m_I$,
and only small variations of the wave function $s_{\{m_I\}}$ between
adjacent lattice points. If we perform the continuum limit
$\gamma\to0$ exactly, then we must simultaneously have $m_I\to\infty$
and $s$ must become a smooth function $s(p^I)$; this is the {\em
pre-classical limit\/} \cite{DynIn}. From a different conceptual point
of view, we can also work with small but non-zero values of $\gamma$,
which is the physically correct way in the case of loop quantum
cosmology where $\gamma=\log(2)/\pi\sqrt{3}$ can be inferred from
calculations of black hole entropy in quantum geometry
\cite{ABCK:LoopEntro,IHEntro}.  In this case we still have to consider
large $m_I$ and small oscillations of the wave functions at Planck
scales for the continuum {\em pre-classical approximation\/} to be
valid. This corresponds to a regime of large volume and small
curvature in the case of gravity.

Concerning the main theme of this paper, the first point of view of a
pre-classical limit is the one usually taken in numerics. When the
differential equation is regarded as fundamental and the difference
equation as an approximation, the limit has to be performed eventually
such that $\gamma$ can be assumed to be as small as necessary for a
particular mathematical result. Thus, mathematical techniques
developed for the numerical analysis of differential equations can be
used when studying the pre-classical limit. When the difference
equation is fundamental, however, $\gamma$ has a physical value which
is non-zero. In such a situation results which rely on the fact that
$\gamma$ can be chosen as small as appropriate, will not apply and new
definitions and theorems have to be developed. In this paper we
present a first approach giving basic definitions and some implications.

%%%%%%%%%%%%%%%%%%%%%%%%%%%%%%%%%%%%%%%%%%%%%%%%%%%%%%%%%%%%%%%%%%%%%%%

\section{Conditions On the Fundamental Difference Equation}
%: \\ \hspace*{1.1cm}Ordinary Difference Equation} 
\label{cond}

To simplify the presentation let us first work in the context of
isotropic loop quantum cosmology \cite{IsoCosmo} so that we have a
single triad component $\hat{p}$ with eigenvalues
$p(m)=\frac{1}{6}\gamma\lP^2m$ and eigenstates $|m\rangle$.
The fundamental difference equation is of the general form (\ref{Gen}) 
with only one summation label,
\begin{equation} \label{DiffEq}
 A_k^m(\gamma)s_{m+k}+A_{k-1}^m(\gamma)s_{m+k-1}+\cdots +
 A_{-k}^m(\gamma) s_{m-k}=0\,,
\end{equation}
where the coefficients satisfy the unitarity conditions for large $m$,
\begin{equation}\label{coefflarge}
 A_{-i}^m\simeq (A_i^m)^*\,.
\end{equation}
The $\gamma$-expansion of the coefficients is such that only one pole
term of order $l=2$ appears, while all other terms have non-negative
power of $\gamma$.

{\em Note:} In practice, the difference equation resulting from a
quantization of the Hamiltonian constraint is not unique. As discussed
in the preceding section, the discrete wave functions $s_m$ are
defined only modulo $m$-dependent phase transformations which also
alters the coefficients of the difference equation. Such a freedom has
to be kept in mind while checking the conditions we will derive for
any particular difference equation. For {\em homogeneous} difference
equations which arise in the isotropic context, the coefficients are
defined only up to common factors. Additionally, for large $m$ in
particular, such common factors do occur but can be canceled. In the
following, the coefficients will be assumed to have no such common
factors.

\subsection{Continuum Approximation}

When a discrete formulation is used at the fundamental level, it has
to reproduce the familiar continuum idealization at large scales or in
semiclassical regimes as a very good approximation.  This presents a
necessary condition for any discrete formulation; in particular a
quantization of geometry has to reproduce the familiar continuum
metric formulation as an approximation when quantum effects can be
expected to be negligible. The latter is realized when the length
scale on which the continuum formulation is used is large compared to
the Planck scale on which quantum geometry is to be used, and
curvature is not significant. In the context of homogeneous cosmology
the corresponding region of phase space has large values of the triads
(and by implication of volume) and small values of the extrinsic
curvature. The intrinsic curvature of a minisuperspace model, by
contrast, is a fixed function on the phase space and thus cannot be
independently required to be small in a semiclassical regime; it can
be large even in regions where the volume is large and the extrinsic
curvature is small, which has to be taken care of properly.

For gravity, the familiar Wheeler--DeWitt quantization of homogeneous
models \cite{DeWitt,Misner} is obtained using a Schr\"odinger
quantization of the continuous geometry variables, i.e.\ the spatial
metric and the extrinsic curvature. In this quantization, the
fundamental formulation is still continuous: wave functions are
functions of continuous labels $p$ and the constraint equation leads
to the Wheeler--DeWitt {\em differential} equation.  Since the
Wheeler--DeWitt quantization already uses continuous geometric
variables, it is convenient (but not necessary since, after all, the
Wheeler--DeWitt quantization is not well-defined for the full theory)
to require that the wave functions of this quantization together with
the Wheeler--DeWitt equation they satisfy should be reproduced as a
suitable {\em limit} or at least as the leading behavior in an {\em
approximation} from some solution(s) of the fundamental difference
equation.

The indication of the regime of semiclassical behavior and the
justification for matching with the Wheeler--DeWitt quantization can
be further seen by observing that in a triad representation, the basic
difference operators of the discrete formulation are given by
$\sin(\Case{1}{2}c)$ as in (\ref{sin}), while the basic differential
operators of a Wheeler--DeWitt quantization are given by
$\hat{c}=i\gamma\lP^2\partial/\partial p$. Thus, if $c$ can be assumed
to be small, we in fact have $\sin(\Case{1}{2}c) \sim\Case{1}{2}
\hat{c}$ and the difference operators immediately yield differential
operators for small $c$. This assumption, however, is justified only
in the simplest cosmological models, the Bianchi I model and its
isotropic sub-models in which the intrinsic curvature, represented by
the spin connection $\Gamma$, is zero and $c$ is essentially the
extrinsic curvature. Otherwise, $c$ also has a contribution from
intrinsic curvature and the relation to the continuum formulation is
less direct; its proper treatment is discussed in
\cite{Closed,HomSpin}.

The condition of large volume is easier to implement since the large
volume regime naturally corresponds to large eigenvalues of the triad
operator, $m \gg 1$. If finite increments of the label $m$ appearing
in the difference equation have only small effects on the wave
function (which also depends on the curvature) then one can expect
difference operators acting on sufficiently slowly varying functions
of $m$ to be well approximated by differential operators. In this way,
one can conceive the possibility of recovering the Wheeler--DeWitt
quantization. Thus, the basic idea is to see if there are
`sufficiently slowly varying' discrete wave functions, $s_m$ which can
be interpolated by `sufficiently slowly varying' solutions of the
Wheeler--DeWitt equation. The requirement that the Wheeler--DeWitt
quantization be obtained as a good approximation translates into the
requirement that the fundamental difference equation admit at least
one such slowly varying solution.  We will now make these statements
precise and deduce their implications.

\subsubsection{Preliminaries}

In the following, the {\em domain of validity} of a continuum approximation
is specified by $1 \ll M_{\text{min}} \ll M_{\text{max}}$ and $m \in 
(M_{\text{min}}, M_{\text{max}})$. For $m_0$ in the domain of validity,
we will specify its $\Delta m$-neighborhood as the interval $[m_0, m_0
+ \Delta m]$. Statements valid over $\Delta m$-neighborhoods will be
termed as {\em local} statements. Typically, $1 \ll \Delta m \ll m_0$
will be assumed. In some situations the domain of validity can extend
over an infinite range of $m$.

The coefficients in the difference equation depend on $\gamma$ and
$m$.  The $\gamma$ dependence is such that $A^m_i =
\gamma^{-2}(A^{(-2)})^m_i + (A^{(0)})^m_i + \gamma (A^{(1)})^m_i +
\cdots $. Note that there is a term of order $\gamma^{-2}$ while there
is no $\gamma^{-1}$ term in the coefficients, as discussed in Section
\ref{s:LQC}. All the $(A^{(r)})_i^m$ also depend on $m$ which,
in a $\Delta m$ neighborhood of some $m_0$ in the domain
of validity, can be expanded as: $(A^{(r)})_i^{m_0 + \delta m}
\simeq A^{(r)}_i + \frac{\delta m}{m_0} B^{(r)}_i + \cdots \ , \ 0 \le
\delta m \le \Delta m$ , where
$A^{(r)}_i, B^{(r)}_i$ are constants depending on the fixed $m_0$ but
independent of $\delta m$.

A few preliminary definitions: A function $\psi$ of a continuous
variable $p$ is said to be {\em locally slowly varying around $p_0$}
if $\psi(p)$ can be Taylor expanded about $p_0$ with succeeding terms
smaller than the preceding terms for $p \in [p_0, p_0 + \Delta p]$. A
bound on the ratios can be used to quantify degree of slow variation
but we will not need it. We will typically need terms only up to
second order. Correspondingly, a sequence $s_m$ is said to be {\em
locally slowly varying around $m_0$} if it can be obtained via $s_m :=
\psi(p(m))$ where $p(m) = \frac{1}{6}\gamma \lP^2 m$, $\psi(p)$ is
locally slowly varying. In other words, a sequence is slowly varying if it
can be interpolated by a slowly varying function, which, of course, will
not be unique.

Notice that the definition of the variable $p(m)$ contains an explicit
$\gamma$ dependence. Thus, even if a slowly varying $\psi(p)$ has no
$\gamma$ dependence, the corresponding slowly varying sequence does
depend on $\gamma$ and is in fact {\em Taylor expandable in}
$\gamma$. Conversely, if a slowly varying sequence has a specific type
of $\gamma$ dependence (because it is a solution of the difference
equation containing $\gamma$, for instance), the corresponding
interpolating function can be independent of $\gamma$.

Thus, a slowly varying function and a slowly varying sequence can 
be written as 
\begin{eqnarray}
\psi(p_0 + \delta p) & \simeq & \psi(p_0) + \delta p \ \psi'(p_0)
 + \frac{1}{2}\delta p^2 \psi''(p_0) + \cdots \ , \label{SlowVar}\\
s_{m_0 + \delta m} & \simeq & \psi(p(m_0)) + (\gamma \lP^2/6) \,
\delta m \ \psi'(p(m_0)) +\frac{1}{2}(\gamma\lP^2/6)^2 \,\delta
m^2\,\psi''(p(m_0))+ \cdots \label{Interpol}
\end{eqnarray}

Let us assume now that the fundamental equation admits solutions which
are locally slowly varying around some $m_0$ in the domain of validity.

\subsubsection{Pre-classicality}

Using the $\gamma$-expansions of the coefficients and substituting
(\ref{Interpol}) for a locally slowly varying $s_{m + i}$ in 
the fundamental difference equation (\ref{DiffEq}) leads to,
\begin{eqnarray}
\sum_{i = -k}^k \left( 
\gamma^{-2}(A^{(-2)})^m_i + (A^{(0)})^m_i + \gamma (A^{(1)})^m_i + \cdots 
\right) \mbox{\hspace{4.0cm}} & & \nonumber \\
\mbox{\hspace{0.0cm}} \times\left( 
\psi(p(m)) + (\gamma \lP^2/6)\,i\, \frac{\md \psi}{\md p} +
\frac{1}{2}(\gamma\lP^2/6)^2\,i^2\, \frac{\md^2 \psi}{\md
p^2} + \cdots
\right) & = & 0
\end{eqnarray}

It is clear that we can get a {\em differential equation} for
$\psi(p)$ in this manner for {\em every locally slowly varying
solution} $s_m$ of the fundamental equation.

Essential for a continuum approximation is not just to get a
differential equation but that a differential equation so obtained
should be {\em independent of $\gamma$}. The $\gamma$ independence
condition is necessary since by definition $\gamma$ controls the
discreteness such that the continuum variables are obtained for
$\gamma \to 0$. For instance, in the case of gravity the differential
equation is expected to be the standard Wheeler--DeWitt equation which
knows nothing about $\gamma$.

It follows immediately that $\psi(p)$ must satisfy equation(s) obtained
by equating coefficients of powers of $\gamma$ to zero. There are, of
course, infinitely many such equations. 

{}From any of these equations, one can derive a `Hamilton--Jacobi'
equation for the phase of the wave function in the usual manner and
read off {\em a classical Hamiltonian constraint}. Only one of these
equations reproduces the correct classical Hamiltonian constraint. In
view of the small value of $\gamma$, this equation satisfied by
$\psi(p)$ is to be considered as the {\em leading approximation} and
consequently, the equations corresponding to lower powers of $\gamma$
must be identically zero {\em without} any conditions on $\psi(p)$.

Thus we define:

\begin{Def}
 A locally slowly varying solution around $m_0$ of the fundamental
 difference equation is said to be {\em locally pre-classical around
 $m_0$} if there is an interpolating slowly varying function $\psi(p)$
 which solves a differential equation that has no dependence on
 $\gamma$ and has the correct classical limit.
\end{Def}

Thus, the $\gamma$ independence condition together with the correct
{\em classical limit}, implies that the $o(\gamma^{-2}),
o(\gamma^{-1})$ terms must be zero identically and the $o(\gamma^0)$
term must contain derivatives of $\psi(p)$. Furthermore, in a $\Delta
m$-neighborhood of $m_0$ we can use $(A^{(r)})^m_i \simeq A^{(r)}_i$
which also removes the explicit $m$ dependence. This leads to:
\begin{equation} \label{CoeffCond}
\sum_{i = -k}^k A^{(-2)}_i = 0 = \sum_{i = -k}^k i A^{(-2)}_i \ ,
~~{\text{and}}~~
\sum_{i = -k}^k A^{(-2)}_i \ i^2  \neq  0 
~~~~~ \forall \ m \in (m_0, m_0 + \Delta m) \ ,
\end{equation}
and we obtain, to the leading order in $\gamma$ and $\frac{\Delta m}{m_0}$, 
the {\em approximating differential equation} as
\begin{equation}\label{ApproxDiffEq}
\frac{1}{72}\lP^4\left( \sum_{i = -k}^k A^{(-2)}_i \ i^2
\right)\frac{d^2 \psi}{dp^2} + 
\left( \sum_{i = -k}^k A^{(0)}_i \right) \psi(p) + o(\gamma,
\Delta m/m_0)  =  0 \ . 
\end{equation}

The approximating differential equation will precisely match with the
Wheeler--DeWitt equation if the ratio of the coefficients of the two
terms on the left hand side of (\ref{ApproxDiffEq}) matches with those
of the Wheeler--DeWitt equation. This indeed turns out to be the
case. We will refer to the leading order approximating differential
equation as the Wheeler--DeWitt equation. Using the Wheeler--DeWitt
equation for capturing the continuum behavior is thus an
approximation. From the perspective of the discrete solutions, locally
pre-classical solutions are still {\em exact} solutions of the
fundamental equation.

Thus we have seen that the admissibility of even a {\em single} locally
pre-classical solution puts strong restrictions on the coefficients of the 
fundamental equation: 

\begin{Prop} \label{Cond}
 If a difference equation of the form (\ref{DiffEq}) with coefficients
 having a second order pole in the parameter $\gamma$ admits a locally
 pre-classical solution, then the leading order terms of the
 coefficients, $A^{(-2)}_i$, must satisfy (\ref{CoeffCond}).
\end{Prop}

We are now ready to define pre-classical solutions.
\begin{Def}[Pre-Classicality] \label{PreClass}
 A solution of the fundamental equation is said to be {\em
 pre-classical} if it is locally pre-classical around every $m_0$ in
 the domain of validity.
\end{Def}

The requirement of the existence of a pre-classical solution implies that
all the local statements above, in particular the conditions in
(\ref{CoeffCond}) must hold throughout the domain of validity.

Notice that these conditions are homogeneous in the coefficients. For
the equations of the isotropic homogeneous models, there are only {\em
three} non-zero coefficients $A^{(-2)}_i$ and the two conditions of
(\ref{CoeffCond}) fix these up to an over-all scaling. It is the
over-all scaling that contains the $m_0$ dependence. The remaining
ratios being independent of $m_0$ are the same for local neighborhoods
around any $m_0$ in the domain of validity.

To summarize: We have reformulated the definition of a pre-classical
solution {\em without} reference to any pre-classical limit
\cite{DynIn}.  It is more realistic since it works with small but
non-zero $\gamma$ and large but not infinite $m$, captures the
essential ideas of getting a continuum description as an approximation
in terms of solutions of the Wheeler--DeWitt equation and directly
gives restrictive conditions (\ref{CoeffCond}) on the coefficients of
the fundamental difference equation. One can construct {\em
approximate} pre-classical solutions from (slowly varying) solutions
of the Wheeler--DeWitt equation as outlined in \cite{DynIn}.

\subsection{Local Stability}\label{LocStab}

We now address the issues of the existence of pre-classical solutions,
the number of independent pre-classical solutions and the construction
of approximate pre-classical solutions. We will work exclusively
within the discrete formulation and assume that the coefficients
of the fundamental equation satisfy the conditions of
(\ref{CoeffCond}) which are necessary conditions for the existence of
a pre-classical solution.  As a consequence, we will construct local
approximations and use them to argue about the existence of
pre-classical solutions.

\subsubsection{Approximate pre-classicality}

Consider locally slowly varying solutions (pre-classical ones being a
special case) of the fundamental equation. We know that such solutions
are Taylor expandable in $\gamma$ : $s_m = s^{(0)}_m + \gamma \
s^{(1)}_m + \gamma^2 \ s^{(2)}_m + \cdots $. Substituting in
(\ref{DiffEq}) and using the fact that the coefficients are locally
almost constant gives,
\begin{eqnarray}
0 & = & 
\gamma^{-2}\left( \sum_{i = -k}^k A^{(-2)}_i s^{(0)}_{m + i} \right)
+
\gamma^{-1}\left( \sum_{i = -k}^k A^{(-2)}_i s^{(1)}_{m + i} \right)
\nonumber \\
& & + \gamma^{0}\left( \sum_{i = -k}^k A^{(-2)}_i s^{(2)}_{m + i} 
+
\sum_{i = -k}^k A^{(0)}_i s^{(0)}_{m + i} \right)
+ o(\gamma)
\end{eqnarray}
It is evident now that a locally pre-classical solution can be
constructed as a power series in $\gamma$ by {\em choosing} 
$s^{(0)}_m , \ s^{(1)}_m$ to be solutions of 
\begin{equation}\label{LeadingEq}
\sum_{i = -k}^k A^{(-2)}_i s_{m + i} = 0 \ ,
\end{equation} 
while $s^{(2)}_m$ is a solution of the non-homogeneous equation
\begin{equation}\label{NonHomoEq}
\sum_{i = -k}^k A^{(-2)}_i s^{(2)}_{m + i} +
\sum_{i = -k}^k A^{(0)}_i s^{(0)}_{m + i}  = 0 \ .
\end{equation}
The higher order terms are to be constructed recursively from solutions of
{\em non-homogeneous} difference equations with constant coefficients.
Truncating the Taylor series to any finite order constitutes an
approximation. Thus we define:

\begin{Def}
 A solution of the form $s_m = s^{(0)}_m + \gamma \ s^{(1)}_m +
 \gamma^2 \ s^{(2)}_m$ with $s^{(r)}_m$ satisfying
 (\ref{LeadingEq},\ref{NonHomoEq}) is said to be a 2nd order
 approximate, locally pre-classical solution or {\em 2nd order local
 approximation} for short.
\end{Def}

Apart from being Taylor expandable in $\gamma$, locally slowly varying
sequences also have the form $s_{m + i} \simeq a_m + i b_m + i^2 c_m +
\cdots $, where $b_m \sim o(\gamma), \ c_m \sim o(\gamma^2)$ etc. as
seen in (\ref{Interpol}). In fact, the equations (\ref{LeadingEq},
\ref{NonHomoEq}) do admit solutions of this form. To see this let us
take the ansatz (for $r = 0, 1, 2 \ $) ,
\begin{equation}
s^{(r)}_{m_0 + \delta m} \simeq a^{(r)}(m_0) + (m_0 + \delta m)
b^{(r)}(m_0) + (m_0 + \delta m)^2 c^{(r)}(m_0) + \cdots ~~,~~ 
0 \le \delta m \le \Delta m \ .
\end{equation}

Substitution in the equations (\ref{LeadingEq}, \ref{NonHomoEq})
implies that $a^{(r)}(m_0)$, $b^{(r)}(m_0)$ are arbitrary,
$c^{(r)}(m_0) = 0$ for $r = 0, 1$ and $c^{(2)}(m_0)$ is determined in
terms of the $a^{(0)}(m_0)$, $b^{(0)}(m_0)$ and coefficients of the
difference equations.  The conditions (\ref{CoeffCond}) are crucial
for these statements. We can {\em choose} the $a^{(r)}(m_0),
b^{(r)}(m_0)$ coefficients so that $s_{m_0 + \delta m}$ is obtained as
a Taylor series in $\gamma \delta m$. Explicitly the choices are:
$b^{(0)}(m_0) = 0, \ a^{(1)}(m_0) = - m_0 b^{(1)}(m_0), \ a^{(2)}(m_0)
= -m_0^2 c^{(2)}(m_0), \ b^{(2)}(m_0) = - 2 m_0 c^{(2)}(m_0)$.  This
leaves us with {\em two} free parameters, $a^{(0)}(m_0), \
b^{(1)}(m_0)$. We have thus constructed 2nd order local approximations
(in a $\Delta m$ neighborhood of $m_0$ and to order $\gamma^2$) which
are parameterized by two constants and given by,
\begin{equation} \label{LocalSoln}
s_{m_0 + \delta m} = a^{(0)}(m_0) + \gamma \delta m b^{(1)}(m_0) +
\gamma^2 \delta m^2 c^{(2)}(m_0) \ .
\end{equation}

The pre-classicality properties of these solutions are obvious. It is
also clear that the number of independent, exact pre-classical
solutions can not be more than {\em two} since their local
approximations are parameterized by two parameters\footnote{All exact
solutions of the fundamental equation have to satisfy the consistency
condition following from the singularity avoidance mechanism
\cite{Sing,DynIn}.  Given a local approximation (for $m \in [m_0, m_0 +
\Delta m]$), we can evolve it backward by the exact fundamental
difference equation. The consistency condition then will lead to a
relation between $a^{(0)}(m_0), b^{(1)}(m_0)$ so that the solution is
determined uniquely up to an over-all scaling (in the vacuum isotropic
case). Such a solution is by construction, exact up to $m_0$ and
approximates a locally pre-classical solution up to $m_0 + \Delta m$.
The number of independent solutions is thus different from the
Wheeler--DeWitt equation, but only after taking into account the
behavior in the Planck regime.} (this is true even if we go to higher
orders in $\gamma$).

We can construct such two parameter families of approximate, locally
pre-classical solution around various $m_j$ in the domain of
validity. These should somehow be tied together to constitute an
approximation to a single (common) exact pre-classical solution, as
discussed in what follows.

\subsubsection{Construction of approximate, pre-classical solutions}
Now we give a procedure to construct a solution which will approximate
a pre-classical solution in various local neighborhoods. To do so,
we divide the domain of validity into several, non-overlapping local
neighborhoods around $m_0 := M_{\text{min}}, m_1, \cdots \ , m_N :=
M_{\text{max}}$ within each of which we have the constant coefficients
$A^{(-2)}_i (m_j)$ satisfying the conditions (\ref{CoeffCond}). We can
construct local approximations in each of these local neighborhoods as
outlined above. These local approximations now have to be `matched'
across the boundaries of the adjacent neighborhoods.

Both the exact (\ref{DiffEq}) and the approximated (\ref{LeadingEq},
\ref{NonHomoEq}) evolution equations are of order $2 k$, and to
determine any particular solution we need $2k$ consecutive $s_m$'s to
be specified as {\em initial conditions}. We will generically refer to
the set $\{s_m, s_{m + 1}, \cdots, s_{m + 2 k -1} \}$ as {\em initial
data at $m$}. In each neighborhood, the local approximations induce
initial data of the form
\begin{equation} \label{InitData}
s_{m_j + i}  =  a^{(0)}(m_j) + \gamma i b^{(1)}(m_j)  + \gamma^2 i^2
c^{(2)}(m_j) + \cdots ~, ~~~i = 0, 1, \cdots , 2k -1\ . 
\end{equation}
Likewise, we also have the $2k$ consecutive values from $s_{m_j + \Delta
m_j - 2k}$ as,
\begin{equation} \label{FinalData}
s_{m_j +\Delta m_j - 2 k + i} =  a^{(0)}(m_j) + \gamma (\Delta m_j -2 k + i)
b^{(1)}(m_j)  + \gamma^2 (\Delta m_j -2 k + i)^2
c^{(2)}(m_j) + \cdots ~. 
\end{equation}

The initial data at $m_{j + 1} = m_j + \Delta m_j$ however can not be
arbitrary since the exact equation holds across the neighborhoods. We
can still use the constant coefficients to connect these values,
however the constant coefficients will now involve a mixture of
coefficients from both the adjacent neighborhoods and this mixture
need not satisfy the conditions (\ref{CoeffCond}). Equivalently, the
initial data so constructed at $m_j + \Delta m_j$ need not equal the
data induced from the local approximation in the neighborhood $[m_{j +
1}, m_{j + 1} + \Delta m_{j + 1}]$. We will refer to the former as the
{\em extended initial data}, the latter as the {\em local data} and
the corresponding solutions of (\ref{LeadingEq}, \ref{NonHomoEq}) as
{\em extended solution} and {\em local solution}. Note that the
initial data consist of $2k$ values while we have a freedom of only
{\em two} parameters in the local data.  Since usually $2k > 2$, the
two data sets are generically distinct and so are the two
solutions. In particular, the extended solution {\em cannot} be a
local continuum approximation. Using a local solution instead of an
extended solution thus introduces errors which can accumulate with
subsequent extensions to cover the full domain of validity. The error
accumulation can even jeopardize the existence of a pre-classical
solution itself. This is prevented by requiring `stability' for the
approximated evolution (\ref{LeadingEq}, \ref{NonHomoEq}).

To arrive at a suitable notion of stability for (\ref{LeadingEq}), let us
concentrate on the equation:
\begin{equation}\label{ApproxEvol}
\sum_{i = -k}^{k} A_{i} s_{m + i} ~ = ~ 0 ~~,~~ \forall ~ m \in 
[m_j + k, m_j + \Delta m - k]
\end{equation}
where $A_i$ are constants which are assumed to satisfy the conditions
(\ref{CoeffCond}). With the identification $A_i := A^{(-2)}_i$, this is
the same as (\ref{LeadingEq}) and has already been referred to as the {\em 
locally approximated evolution}. 

It is convenient to denote the initial data by a vector in
$\mathbb{C}^{2k}$, $(\vec{S}(m_j))^i := s_{m_j + i}~, ~ i = 0, 1,
\cdots, 2k -1$. On the space of initial data, we have the natural
Euclidean norm $\mathbb{C}^{2k}, ||\vec{S}(m_j)||^2 := \sum_{i = 0}^{2
k -1} |s_{m_j + i}|^2$.

Consider now the approximated evolution of two `nearby initial data'. Let 
$\vec{S}(m_j)$ be fixed initial data and consider other arbitrary initial 
data $\vec{S}'(m_j)$  such that $||\vec{S}'(m_j) - \vec{S}(m_j)|| =: 
\delta_{m_j} \le \Delta $. These data will generate two solutions $s'_m, 
s_m$ whose difference, $\delta s_m$, will also satisfy the approximated 
evolution equation. We would like to see how much the two solutions
differ at $m_j + \Delta m - 2 k$. 

The general solutions of difference equations of constant coefficients
are given by linear combinations of elementary solutions of the form
$(z_a)^m m^{r_a}$ where $z_a$ is a root of the {\em characteristic
polynomial}, $\sum_i A_i z^{i}$, with multiplicity $d_a$ and $0 \le
r_a \le d_a - 1$ \cite{elaydi}. Notice that (\ref{CoeffCond}) implies
that $z = 1$ is a root with multiplicity 2. Clearly, $\delta s_m =
\sum_a \sum_{r_a = 0}^{d_a} \delta C_{a, r_a} \ (z_a)^m m^{r_a}$ where
$a$ ranges over the distinct characteristic roots. Since
$\vec{S}'(m_j)$ is arbitrary, consider a $\delta s_m$ such that
$\delta C_{a, r_a}$ is non-zero only for a particular elementary
solution. Then we have,
\begin{equation} \label{Growth}
\delta_{m_j + \Delta m -2 k} \simeq 
|z_a|^{\Delta m -2 k} \left(1 + r_a \frac{\Delta m - 2 k}{m_j} + \cdots 
\right) \ \delta_{m_j}  \ .
\end{equation}

Clearly, if there is a characteristic root with absolute value greater
than one, then the two solutions deviate exponentially. Taking the data 
$\vec{S}_{m_j}$ to be local data and $\vec{S}'_{m_j}$ to be the extended 
data,  we see that even if the two data are within a $\Delta$ neighborhood 
initially, they could evolve far apart if $|z_a| > 1$. This leads us to
the definition:
\begin{Def}
 The approximated evolution (\ref{ApproxEvol}) is said to be {\em
 stable} if $\delta_{m_j + i } \le \Delta( 1 + o(\frac{\Delta
 m_j}{m_j}) )\ \forall \ 0 \le i \le \Delta m - 2k$ for all
 $\delta_{m_j} \le \Delta$.
\end{Def}

A stable evolution then guarantees that the extended solution will be nearby
the local solution i.e. $|\delta s_m| \le \Delta$. It implies that {\em all}
characteristic roots must have absolute value less than or equal to one. The 
coefficients of the difference equation also satisfy the complex conjugation
property (\ref{coefflarge}) which implies that if $z$ is a root, so is
$(z^*)^{-1}$. Combined with the stability requirement this implies 

\begin{Prop} \label{Roots}
 The approximated evolution of a difference equation (\ref{DiffEq})
 with coefficients satisfying the unitarity conditions
 (\ref{coefflarge}) is stable if and only if all characteristic roots
 have absolute value one.
\end{Prop}

As noted above, in the `first neighborhood' ($[m_0, m_0 + \Delta
m_0]$), we are left with a single free parameter, say $a^{(0)}(m_0)$,
due to the consistency condition. In the next neighborhood we have two
free parameters, $a^{(0)}(m_1), b^{(1)}(m_1)$. We can fix these by
requiring that the extended and the local data at $m_1$ differ by the
least amount eg. by a least square fit. This determines the free
parameters and also determines $\delta_{m_1}$.  We can take $\Delta =
\delta_{m_1}$. If the locally approximated evolution is stable in the
second neighborhood, then the local solution in the second
neighborhood is guaranteed to be within $\delta_{m_1}$ of the solution
extended from the first neighborhood. If locally approximated
evolutions are stable for all the neighborhoods, then we can continue
our procedure and construct a sequence $s_m$ with the property that it
is an exact solution up to $m_0$ and is an approximate, locally
pre-classical solution in every neighborhood of the chosen partition
of the domain of validity. It is thus essential for the construction
of an {\em approximate, piece-wise, pre-classical solution} that all
the locally approximated evolutions be stable. This leads us to
define:

\begin{Def}[Local Stability] \label{DefLocStab}
 The fundamental difference equation is said to be locally stable if the
 locally approximated evolution in any local neighborhood around any
 $m_0$ in the domain of validity, is stable. In particular this implies
 that the coefficients of the fundamental equation satisfying
 (\ref{CoeffCond}) have to satisfy the further condition that their
 corresponding characteristic polynomials have all roots with absolute
 value one.
\end{Def}

In other words,
\begin{equation}\label{StabCond}
\sum_{i = -k}^k A^{(-2)}_i(m_0) z^i(m_0) ~ = ~ 0 ~~ {\text{implies}} ~~
|z(m_0)| = 1 , ~ \forall ~ m_0 ~ \in ~ [M_{\text{min}}, M_{\text{max}}] \ .
\end{equation}

Note that the construction of an approximate pre-classical solution
depends on the choice of $M_{\text{min}}$ (up to which exact evolution
is used) and also on the partitioning of the domain of validity. Hence
such approximate pre-classical solutions are not unique. This is also
true of {\em exact pre-classical solutions}. Since one does not know
what initial data at $m = 0$ would characterize a pre-classical
solution, we can choose the local data at any $m_0 \in
[M_{\text{min}}, M_{\text{max}}]$ and evolve it exactly (in both
directions) to generate {\em an} exact pre-classical solution. However
the fact that approximate locally pre-classical solutions match with
pre-classical solutions to order $\gamma^2$ and the property of local
stability guarantees that a pre-classical solution defined from any
local data and any of the approximate pre-classical solutions
constructed, will always be `nearby'. For instance, if we denote by
$\epsilon$ the largest of the $\delta_j/\sqrt{2 k}$ over all the
neighborhoods and all partitions, then we can expect any of the
approximations or any of the exact pre-classical solutions to satisfy
pairwise
\begin{equation}
|s_m - s'_m | ~ < ~ \epsilon ~~ \forall ~ m \in 
[M_{\text{min}}, M_{\text{max}}]
\end{equation}
Thus, our reformulated definition of pre-classicality, allows us to
construct approximate pre-classical solutions to within
$\epsilon$-tolerance.

Several remarks are in order.

(1) The requirement of admissibility of a pre-classical solution and
of local stability are both conceptually and logically independent. It
is the requirement of admissibility of a pre-classical solution that
implies the local constancy of the coefficients $(A^{(-2)})^m_i$
together with the condition (\ref{CoeffCond}) on these
constants. These are explicitly picked out by the requirement of
getting an approximating differential equation (Wheeler--DeWitt
equation) which is independent of $\gamma, \ m$ explicitly. As a
by-product, this requirement provides us with locally approximated
evolutions and also approximate, locally pre-classical solutions which
are parameterized by two free parameters. The local stability on the
other hand is concerned with the construction of an approximate
pre-classical solution by `joining' together the various local
approximations. Because the order of the fundamental evolution
equation is always larger than two in loop quantum cosmology, the
extension of local solutions introduces perturbations of local data
which only stay small if the approximated evolution is stable. By
contrast, a similarly locally approximated {\em differential} equation
does not need stability for an extension since the initial data are
local and we have exactly the same number of parameters as needed for
matching.

(2) Although local stability is needed for approximate pre-classical
solutions, the condition itself refers to the characteristic roots
which are {\em independent of} the pre-classical solution. The
stability property of the approximated evolution generically ensures
that, compared to a given solution, nearby `initial' data evolve into
nearby `final' data. If any of the non-pre-classical solution can also
be approximated locally, then the extensions of its local
approximations will involve similar perturbations of initial
data. Note however that non-pre-classical solutions would generically
be parameterized by more parameters ($2 k - 2$) which are available
for performing an exact matching and stability will not be a concern
there (and not be necessary from a physical point of view).

(3) The implication of stability for the roots is independent of their
multiplicities (less than or equal to $2k -2$) which hardly affect the
growth of deviations. The dangerous exponential growth is prevented by
stability.  Strictly speaking, the definition of stability allows
roots to have absolute values to be $1 + o((m_j)^{-1})$. With finite
domains for local neighborhoods, one {\em cannot} hope to derive
strict equalities. However, even without strict equalities, stability
requirement puts severe constraints on the coefficients.

(4) It is important to note that the need for stability arises out of
matching local approximations in the adjacent neighborhoods which
always introduces differences between extended initial data and local
initial data. It only matters that the local approximations are
obtained as solutions of an initial value problem (i.e. evolution
equations). Whether the evolution is via a homogeneous difference
equation or via a non-homogeneous one, at best controls the difference
between the two initial data sets. For the evolution of {\em
  differences} of solutions, it is {\em only} the homogeneous part of
the evolution that matters. This observation will be particularly relevant
for fundamental equations which are partial difference equations.

\subsubsection{Further Implications for Loop Quantum Cosmology}

%In the context of loop quantum cosmology, if there is a regime in
%which the local approximation extends to infinite volume, local
%stability ensures that all solutions to the constraint equation can be
%interpreted as distributions on the full kinematical Hilbert space as
%test function space, not just on the cylindrical subspace of finite
%linear combinations of the states $|m\rangle$. If the coefficients of
%a solution grew exponentially at large $m$, its evaluation in a
%normalizable state would be ill-defined, prohibiting a distributional
%interpretation. The condition of solutions being distributions on the
%kinematical Hilbert space is, however, less general than local
%stability since the latter can be imposed in any region of large, but
%not necessarily infinite volume. Note also that there can be
%large-volume effects, mainly in the presence of a positive
%cosmological constant, which lead to exponentially growing wave
%functions at large volume even if it is a classically allowed
%region. In this case, however, the continuum approximation would not
%even be valid at too large a volume where the validity of the model
%breaks down due to infrared problems (the oscillation frequency in the
%minisuperspace model is increased by a volume factor which pushes the
%oscillation length below the discreteness scale; see also
%\cite{IsoCosmo}).

Local stability provides a strict selection criterion which
allows us to reduce the number of choices for a quantization of the
Hamiltonian constraint in homogeneous models. This is because for a
random polynomial with coefficients satisfying (\ref{coefflarge}) it
is very unlikely that all roots will have norm one. Also a
quantization of the constraint using holonomy operators in some form
will violate this condition in most cases. This is in particular
helpful because models with non-zero intrinsic curvature require an
additional input which is not present in the full theory and thus are
less close to full quantum gravity. Only the Bianchi I model with
vanishing intrinsic curvature does not require additional steps, and
its quantization which has considerably less freedom turns out to be
locally stable automatically.

Local stability was motivated by semiclassical considerations, and it
has to be fulfilled only in semiclassical regimes. In the Planck
regime it can be violated, and it usually is in the presence of matter
or other effects which lead to large curvature. In such a situation it
is expected that contributions to the wave function with Planck scale
oscillations become dominant, leading to large differences between the
discrete and a continuum formulation. Usually, an unbounded growth of
the strongly oscillating solutions when one approaches the classical
singularity is suppressed by a consistency condition which follows
from the constraint equation \cite{DynIn} as can, e.g., be seen from
the discussion in \cite{Scalar} for the isotropic model.

To summarize our results for the case of isotropic loop quantum cosmology: Our
requirements on the fundamental difference equation (\ref{DiffEq}) are
(a) the equation admits at least one pre-classical solution and (b) it
is locally stable. The former leads to the conditions
(\ref{CoeffCond}) while the latter requires that all the roots of the
characteristic polynomial of the locally approximated difference
equation with constant coefficients have unit absolute value. The
conditions (\ref{CoeffCond}) also imply that among these roots is the root 
$\lambda = 1 + o(\gamma)$ with multiplicity two and this in turn is
responsible for the two parameter family of approximate, locally
pre-classical solutions. The fact that the multiplicity is two can directly
be traced back to the second order pole in $\gamma$ of the coefficients. 
A generalization to a different pole structure is straightforward.

\section{Partial Difference Equations}

So far we concentrated on difference equations which were ordinary,
homogeneous difference equations. Very often, one encounters {\em
partial difference equations} as in the case of anisotropic, homogeneous
loop quantum cosmology. The rationale and the logic for continuum
approximations, pre-classicality etc remains the same but the details of
course differ. We will concentrate on the differences one encounters and
keep the discussion general. The concrete case of homogeneous
loop quantum cosmology will be used as an example; it is discussed in
more detail in \cite{HomSpin}.

\subsection{Continuum Approximation}

The fundamental partial difference equation is of the form,
\begin{equation}
\sum_{\{i_I\}} A_{\{i_I\}}^{\{m_I\}}(\gamma) s_{\{m_I+i_I\}}=0
~~\text{with}~~
A_{\{i_I\}}^{\{m_I\}}(\gamma)= \sum_{j=-l}^{\infty} \gamma^j
 (A^{(j)})_{\{i_I\}}^{\{m_I\}}
\end{equation}

The locally slowly varying sequence is given in terms of an 
interpolating, locally slowly varying function $\psi(p)$ by 
(recall $p^I(m_I) = \gamma\ell m_I$),
\begin{eqnarray}
s_{\{m_I+\delta m_I\}} & = & s_{\{m_I\}} +
\gamma\ell\sum_J \delta m_J \partial_J\psi(p^I(m_I))+
 \frac{1}{2} \gamma^2\ell^2 \sum_{J,K}\delta m_J\delta m_K
 \partial_J\partial_K \psi(p^I(m_I))+\cdots \ .
\end{eqnarray}

The $\gamma$-independent equations obtained from the fundamental
equation become,
\begin{equation}\label{GammaZero}
 \left(\sum_{\{i_I\}}(A^{(-l)})_{\{i_I\}}^{\{m_I\}}\right)
 \psi(p^I(m_I)) =0\,,
\end{equation}
\begin{equation}\label{GammaOne}
 \left(\sum_{\{i_I\}}(A^{(-l+1)})_{\{i_I\}}^{\{m_I\}}\right)\psi(p^I(m_I))+
 \ell\sum_J
 \left(\sum_{\{i_J\}} (i_JA^{(-l)})_{\{i_I\}}^{\{m_I\}}\right)
 \partial_J \psi(p^I(m_I))=0
\end{equation}
and so on.

Once again one has to appeal to recovery of the correct continuum
equation (analogous to the classical limit) to select the leading order 
partial differential equation. In the case of general homogeneous loop quantum
cosmology ($l = 2$ and no $\gamma^{-1}$ term), this requires the differential
equation to be second order. 

For a non-zero solution, the bracket in eq.(\ref{GammaZero}) must be zero. 
Using (rotational) symmetry conditions for the coefficients, we can
reformulate this condition as
\begin{equation}\label{CondZero}
\sum_{i_I}(A^{(-l)})_{\{i_I\}}^{\{m_I\}}=0
\end{equation}
(summing only over a single $i_I$) for all $I$ and all choices of
$i_J$ for $J\not=I$.

The second equation (\ref{GammaOne}) either gives us a first order
differential equation or leads to further conditions on the
coefficients of the difference equation and so on. {\em Let us assume
for definiteness that the desired continuum equation is of second
order}. Then the additional conditions for the coefficients are,
\begin{equation}\label{CondOne}
 \sum_{\{i_I\}}(A^{(-l+1)})_{\{i_I\}}^{\{m_I\}}=0
\end{equation}
and
\begin{equation}\label{CondTwo}
 \sum_{\{i_I\}}(i_JA^{(-l)})_{\{i_I\}}^{\{m_I\}}=\sum_{i_J}i_J
 \sum_{\{i_I\}_{I\not=J}} (A^{(-l)})_{\{i_I\}}^{\{m_I\}}=0
\end{equation}
for all $J$ (this cannot be split into several conditions by using
symmetry).

For homogeneous loop quantum cosmology, the conditions on the
coefficients are
\begin{eqnarray}\label{HomCoeffCond}
\sum_{\{i_I\}}(A^{(-2)})_{\{i_I\}}^{\{m_I\}} & = & 0\,, \\
\sum_{\{i_I\}}(i_JA^{(-2)})_{\{i_I\}}^{\{m_I\}} & = & \sum_{i_J}i_J
\sum_{\{i_I\}_{I\not=J\}}} (A^{(-2)})_{\{i_I\}}^{\{m_I\}} ~ = ~ 0 \ , 
\end{eqnarray}
and are in fact satisfied for (\ref{HomCosmoCoeff}). In this case, the
approximating partial differential equation is the Wheeler--DeWitt
equation for the model.

The approximating {\em partial differential equation} becomes,
\[
\ell^2\sum_{J,K}\left( \sum_{\{i_I\}} i_J i_K 
(A^{(-l)})^{\{m_I\}}_{\{i_I\}} \right) \partial^2_{J K} \ \psi(p^I) + 
\ell\sum_J \left(\sum_{\{i_J\}} (i_JA^{(-l + 1)})_{\{i_I\}}^{\{m_I\}}\right)
\partial_J \psi(p^I(m_I)) 
\]
\begin{equation}\label{ApproxPDE}
+ 
\left(\sum_{\{i_I\}}(A^{(-l + 2)})_{\{i_I\}}^{\{m_I\}}\right) \psi(p^I)
 =  0 \ .
\end{equation}

Note that the bracket in the first term can be zero for some values of
$J, K$ but it has to be {\em non-zero} for those $J, K$ which will
reproduce the correct continuum equation. The desired continuum
approximation thus gives us the condition (\ref{CondZero}, \ref{CondOne},
\ref{CondTwo}).

\subsection{Local Stability}

The next step is to work with the difference equation in a local
neighborhood and characterize the local {\em continuum} solutions
(analogues of pre-classical solutions). These are Taylor expandable in
$\gamma$ and lead to a hierarchy of partial difference equations. The
$s^{(0)}_{\{m_I\}}$ satisfies a partial difference equation with
constant coefficients while the higher order terms in the Taylor
expansion are determined recursively from {\em non-homogeneous}
partial difference equations with constant coefficients. In a local
neighborhood, we can take a short cut and directly take an ansatz for
a local pre-classical solution as,
\begin{equation}\label{Taylor}
s_{\{(m_0)_I + \delta m_I\}} = a^{(0)}((m_0)_I) + \gamma \sum_{I} \delta m_I
b^{(1)}_I((m_0)_I) + \gamma^2 \sum_{I,J} \delta m_I \delta m_J
c^{(2)}_{IJ}((m_0)_I) +
\cdots ~ .
\end{equation}

Substitution then implies that $a^{(0)}((m_0)_I), b^{(1)}_I((m_0)_I)$
are free parameters analogous to the case of ordinary difference
equation. Now however there are additional possible free parameters, for
example the $c^{(2)}_{IJ}((m_0)_I)$ for those $I, J$ for which the first 
bracket in eqn(\ref{ApproxPDE}) is zero. Similar argument implies that
there could be free parameters at higher orders in the Taylor expansion
(\ref{Taylor}). This is different from the case of ordinary difference 
equations where the higher order coefficients are all determined in
terms of the $a^{(0)}, b^{(1)}$. The local continuum solution is
thus parameterized by potentially infinitely many free parameters.
Extension of a local continuum solution to the adjacent `cells' of
the given local neighborhood will be more complicated.

In the previous section, for formulation and analysis of `local
stability' we concentrated on the {\em locally approximated evolution},
equation (\ref{ApproxEvol}). The analogue of this equation is the equation
satisfied by $s^{(0)}_{\{m_I + i_I\}}$, namely,
\begin{equation}\label{ApproxPDfnE}
\sum_{\{i_I\}}(A^{(-2)})_{\{i_I\}}\ s^{(0)}_{\{m_I + i_I\}}  =  0\,. 
\end{equation}
The $(m_0)_I$ dependence of the coefficients is suppressed.

Unlike ordinary difference equations, this is not an evolution equation
and we do not have a handle on its general solution. We can think of
fixing $m_I + i_I, I \ne 1$ and allowing variation only along the `first
direction' and see if an evolution equation can be arrived at.

Let the lattice be $D$-dimensional. Let us introduce the notation:
\begin{eqnarray}
Z^{i_2, i_3, \cdots, i_D}_{m_1}(m_2, m_3, \cdots, m_D) & := & 
s^{(0)}_{m_1, m_2 + i_2, \cdots, m_D + i_D} \ , \nonumber \\
X^{i_2, i_3, \cdots, i_D} & := & \sum_{i_1} (A^{(-2)})_{\{i_I\}} \
s^{(0)}_{\{m_I + i_I\}} \nonumber \\
& = & \sum_{i_1} (A^{(-2)})_{i_1, i_2, \cdots, i_D} \ 
Z^{i_2, i_3, \cdots, i_D}_{m_1 + i_1 }(m_2, m_3, \cdots, m_D)
\end{eqnarray}

The equation (\ref{ApproxPDfnE}) then can be expressed as,
\begin{equation}
\sum_{i_2,i_3, \cdots, i_D} X^{i_2, i_3, \cdots, i_D} ~ = ~ 0 \,. 
\end{equation}

We can separate a particular set of values $i_2^*, i_3^*, \cdots,
i_D^*$ and view the above equation as an equation for $X^{i_2^*,
i_3^*, \cdots, i_D^*}$. Provided, $X^{i_2^*, i_3^*, \cdots, i_D^*}$,
involves a non-trivial combination of $Z^{i_2^*, i_3^*, \cdots,
i_D^*}_{m_1 + i_1 }(m_2, m_3, \cdots, m_D)$ (i.e. more than one $m_1 +
i_1$ occurring in the sum), we can think of the equation as an
ordinary, non-homogeneous difference equation with respect to $m_1$,
for the $Z^{i_2^*, i_3^*, \cdots, i_D^*}_{m_1 + i_1 } (m_2, m_3,
\cdots, m_D)$. Thus we can interpret the partial difference equation
as a (non-homogeneous) evolution equation for evolution along a line,
parallel to the $m_1$ direction and passing through $m_2 + i_2^*, m_3
+ i_3^*, \cdots, m_D + i_D^*$ in the lattice.

For example, in the context of homogeneous loop quantum cosmology, the
lattice is three dimensional and the indices $i_I$ range from $-2$ to
$+2$. The locally approximated partial difference equation
(\ref{ApproxPDfnE}) turns out to be of the form,
\begin{eqnarray} \label{HomCosEq}
 ~~~d(m_1)\left( s^{(0)}_{m_1, m_2 + 2, m_3 + 2} 
+ s^{(0)}_{m_1, m_2 - 2, m_3 -2} 
- s^{(0)}_{m_1, m_2 + 2, m_3 - 2} 
- s^{(0)}_{m_1, m_2 - 2, m_3 + 2} \right)  & & \nonumber \\
 + ~ d(m_2)\left( s^{(0)}_{m_1 + 2, m_2, m_3 + 2} 
+ s^{(0)}_{m_1 - 2, m_2, m_3 - 2} 
- s^{(0)}_{m_1 - 2, m_2, m_3 + 2} 
- s^{(0)}_{m_1 + 2, m_2, m_3 - 2} \right)  & & \nonumber \\
 + ~ d(m_3)\left( s^{(0)}_{m_1 + 2, m_2 + 2, m_3} 
+ s^{(0)}_{m_1 - 2, m_2 - 2, m_3} 
- s^{(0)}_{m_1 - 2, m_2 + 2, m_3} 
- s^{(0)}_{m_1 + 2, m_2 - 2, m_3} \right) & = & 0 \ , 
\end{eqnarray}
where $d(m) \approx m^{-1}$ for large $m$.

The equation is invariant under permutation of the three directions,
so it is enough to consider `evolution' along the
$`m_1$'-direction. For the $(i_2, i_3)$ set we have the eight
possibilities $(2,2), (2, -2), (-2,2), (-2, -2), (2,0), (-2,0), (0,2),
(0, -2)$ and the corresponding $Z^{i_2, i_3}_{m_1 + i_1}$. The
corresponding $X^{i_2, i_3}$ involve non-trivial combinations of $Z$'s
only for the four sets $(2,0), (-2,0), (0, 2), (0,-2)$.  For example,
$X^{2,0} = d(m_3)(Z^{2,0}_{m_1 + 2} - Z^{2,0}_{m_1 - 2})$ is a
non-trivial combination while $X^{2,2} = d(m_1)Z^{2,2}_{m_1}$ is a
trivial combination. Clearly, the partial difference equation can be
viewed as an ordinary, non-homogeneous difference equation for any of
the four non-trivial combinations. For each of these, the order of the
evolution equation with respect to $m_1$ is 4. This illustrates how
the locally approximated partial difference can be viewed as
(different) non-homogeneous evolution equations for different
combinations, $X$'s.

Matching various local solutions across adjacent cells requires, in
particular, matching along various directions connecting the adjacent
cells. Now the arguments used for the case of ordinary difference
equations can be invoked and issue of local stability becomes relevant.
Note that these arguments are relevant {\em only} for the non-trivial
combinations. For trivial combinations, local solutions in adjacent
cells, across a cell boundary are {\em decoupled} and hence matching
requires only equating the values on the cell boundary.

As observed before, the criteria for local stability depend only on
the equation being an evolution equation and not on its
(non-)homogeneity property. The criterion in terms of the
characteristic roots of the associated {\em homogeneous} equation,
$X^{i^*_2, i^*_3, \cdots, i^*_D} = 0$, then remains the same. This
criterion is of course to be applied to {\em all} the evolution
equations for different sets of $(i_2, i_3, \cdots, i_D)$ and along
each of the directions. These details depend very much on the precise
coefficients $A^{(-2)}_{i_1, i_2, \cdots, i_D}$ and are to be analyzed
on a case by case basis.

For the equation of homogeneous loop quantum cosmology,
eq.(\ref{HomCosEq}), the characteristic roots of the evolution
equations are $\pm 1, \pm i$ which all have absolute value 1 thus
satisfying the criterion of local stability \cite{HomSpin}.

To summarize: The rationale for continuum approximation and local
stability remains the same as in the case of ordinary difference
equation. In particular, with regards to local stability, one still
deals with the locally approximated partial difference equation.
Whenever this equation can be viewed as {\em an evolution} equation,
we require this evolution to be locally stable as defined in the
subsection \ref{LocStab}. Since there could be multiple ways of
viewing the partial difference equation as an evolution equation,
local stability of all these evolutions is required.

\section{Conclusions}

In this paper we have taken the view point that the fundamental
equations of a theory arise from a discrete formulation and are thus
difference equations.  Typically such a fundamental discrete
formulation arises from a quantization procedure which is not free of
inherent ambiguities. A basic requirement on any such discrete
formulation is that a continuum approximation should be available in
suitable regimes. The parameter(s) controlling the discrete structure,
being physical, cannot be taken to zero in arriving at a continuum
approximation. A continuum approximation is then to be derived by
invoking a suitable notion of pre-classicality without using
unphysical, mathematical limits. In the context of isotropic loop
quantum cosmology, we have shown how the definitions of slowly varying
solutions together with $\gamma$-independence of the corresponding
approximating differential equation suffices. Furthermore the
formulation is powerful enough to put constraints on the coefficients
of the fundamental equations as in equation (\ref{CoeffCond}). These
are however necessary conditions for a continuum approximation to be
admissible. The sufficiency can also be demonstrated in a well defined
approximate sense by a constructive procedure provided, in addition,
the condition of local stability (\ref{StabCond}) is satisfied.

More generally, the fundamental difference equation will be a
{\em partial difference equation}. These arise for instance, in the
context of the more general homogeneous models. 
The logic of admissibility of a continuum approximation is
still valid and is seen to lead to similar conditions on the
coefficients. One is then lead to a {\em locally approximated partial
difference equation}. For the favorable case when the locally approximated 
partial difference equation can be viewed as a non-homogeneous, evolution 
equations along some direction for some $s^{(0)}_{\{m_I + i_I\}}$, one
can extend local solutions across neighboring cells in a manner similar
to the case of ordinary difference equation. The issue of stability of
local approximations arises again for exactly the same reasons and leads
to the definition of local stability exactly as before. The local
stability criteria however are to be applied to {\em all} the evolution
equations implied by the locally approximated partial difference
equation. For the fundamental equation of the diagonal, homogeneous
models, these evolution equations turn out to be stable \cite{HomSpin}.

When the fundamental difference equation is derived from a (quantum)
theory, there are usually free parameters or other choices (e.g.,
quantization ambiguities). In such a case the conditions derived here
can serve as powerful restriction criteria which go beyond the usual
requirement that the quantization have the correct classical
limit. Thus, they provide consistency checks for possible quantization
approaches and restrict possible choices within a given approach. In
this light, it is quite remarkable that the loop quantization of the
homogeneous Bianchi I model, which is very close to the quantization
of the full theory and does not allow many choices, turns out to
satisfy our conditions automatically. This can be regarded as a
positive consistency check for the full theory. For other homogeneous
models, which have non-zero intrinsic curvature and are therefore not
as close to the full theory, there are more choices which in
particular allow violations of the conditions. In this case the
conditions serve as strong selection criteria since it is possible to
find quantizations fulfilling them for all those models
\cite{HomSpin}.

%%%%%%%%%%%%%%%%%%%%%%%%%%%%%%%%%%%%%%%%%%%%%%%%%%%%%%%%%%%%%%%%%%%%%%%

\begin{acknowledgments}

We are grateful to K.\ Vandersloot for discussions.
The work of M.~B.\ was supported in part by NSF grant PHY00-90091 and the 
Eberly research funds of Penn State.
\end{acknowledgments}

%%%%%%%%%%%%%%%%%%%%%%%%%%%%%%%%%%%%%%%%%%%%%%%%%%%%%%%%%%%%%%%%%%%%%%%

\appendix*

\section{An Example of an Unstable Quantization}

In this appendix we will focus on the isotropic subclass of the
Bianchi-I and Bianchi-IX models and examine the stability properties
of the evolution equations resulting from the quantization of the
constraint operator as given originally in \cite{IsoCosmo}. Again for
our purposes, it is sufficient to consider only the vacuum case
without the cosmological constant.

Let us recall some basic expressions: An orthonormal basis of states
in the connection representation is given by $\langle c | n \rangle$
analogously to (\ref{cm}), which also forms an eigenbasis of the
volume operator. In the isotropic case the convention $V_{\frac{|n|
-1}{2}}=(\frac{1}{6}\gamma\lP^2)^{\frac{3}{2}} ~ \sqrt{ (|n|
-1)~|n|~(|n| + 1)}$ has been used. As in the full theory \cite{QSDI},
the constraint $\hat{H} = - \hat{H}_E ~ + ~ \hat{P}$ consists of two
parts which can be quantized separately (in the main text we quantized
the full constraint directly which cannot be done in the full theory):
\[
 \hat{H}_E = 4 i (\gamma \kappa \lP^2 )^{-1} \sum_{I,J,K}
\epsilon^{IJK} \tr( h_I h_J h_I^{-1} h_J^{-1} h^{-1}_{[I,J]} h_K
 [h^{-1}_K, \hat{V}] )
\]
where $h_{[I,J]} := \prod_K (h_K)^{C_{IJ}^K}$ with the structure
constants $C_{IJ}^K =\epsilon_{IJK}$ for the closed isotropic model
(Bianchi IX) and zero for the flat model (Bianchi I). The second part
of the constraint operator is obtained from the first one by
\[
 \hat{P} = -8 i (1 + \gamma^{-2}) \kappa^{-1}(\gamma\lP^2)^{-3}
 \sum_{I,J,K} \epsilon^{IJK} \tr ( h_I [h^{-1}_I, \hat{K} ] h_J
 [h^{-1}_J, \hat{K} ] h_K [h^{-1}_K, \hat{V} ] )
\]
with $\hat{K} = - \frac{i}{2} \gamma^{-2}\hbar^{-1} [ \hat{H}_E, \hat{V} ]
~ \widehat{\mbox{sgn(det $a^i_I$ )}}\,.$

Upon transforming to the triad representation by $|s\rangle:= \sum_{n
\in \mathbb Z} S_n |n\rangle$, the constraint equation $\hat{H}
|s\rangle = 0$ gets translated to a difference equation for the
coefficients $S_n$ which are just complex numbers in the vacuum
case. The order of this equation is 16 for Bianchi-I and 20 for
Bianchi-IX. Since the coefficient of $S_0$ is always zero in the
factor ordering chosen for the constraint operator, one can ignore the
sgn factors which enter through the definition of the $\hat{K}$
operator i.e. equivalently absorb them in the $S_n$'s. Explicitly the
equations are:

For Bianchi-I: 
\begin{eqnarray}
0 & ~=~ & -~ \Big( V_{\frac{|n + 4|}{2}} - V_{\frac{|n + 4| - 2}{2}} \Big) 
S_{n + 4} 
 - \Big( V_{\frac{|n - 4|}{2}} - V_{\frac{|n - 4| - 2}{2}} \Big) S_{n - 4} 
\\ 
& &  + 2 \Big( V_{\frac{|n|}{2}} - V_{\frac{|n| - 2}{2}} \Big) S_{n} 
 ~ + ~ \Big(\frac{1 + \gamma^{-2}}{4}\Big) \Big\{ ~
\Big( V_{\frac{|n + 8|}{2}} - V_{\frac{|n + 8| - 2}{2}} 
\Big) \Big(k^+_{n + 8} k^+_{n + 4} \Big) S_{n + 8} \nonumber\\
& & +~ \Big( V_{\frac{|n - 8|}{2}} - V_{\frac{|n - 8| - 2}{2}}
 \Big) \Big(k^-_{n - 8} k^-_{n - 4} \Big) S_{n - 8} 
- \frac{1}{2} \Big( V_{\frac{|n|}{2}} - V_{\frac{|n| - 2}{2}} \Big) 
\Big( k^-_{n} k^+_{n + 4}  + k^+_n k^-_{n - 4} \Big) S_{n} \Big\}
\nonumber
\end{eqnarray}

For Bianchi-IX: 
\begin{eqnarray}
0 & ~~=~~ & + 
\Big( V_{\frac{|n + 5|}{2}} - V_{\frac{|n + 5| - 2}{2}} \Big)
\Big(\frac{1 - i}{2} \Big)  S_{n + 5} + 
\Big( V_{\frac{|n - 5|}{2}} - V_{\frac{|n - 5| - 2}{2}} \Big)
\Big(\frac{1 + i}{2} \Big)  S_{n - 5} \nonumber \\
& & + \Big( V_{\frac{|n + 3|}{2}} - V_{\frac{|n + 3| - 2}{2}} \Big)
\Big(\frac{1 + 5i}{2} \Big)  S_{n + 3} + 
\Big( V_{\frac{|n - 3|}{2}} - V_{\frac{|n - 3| - 2}{2}} \Big)
\Big(\frac{1 - 5i}{2} \Big)  S_{n - 3}  \nonumber\\
& & - \Big( V_{\frac{|n + 1|}{2}} - V_{\frac{|n + 1| - 2}{2}} \Big)
\Big(1 + i\Big)  S_{n + 1} - 
\Big( V_{\frac{|n - 1|}{2}} - V_{\frac{|n - 1| - 2}{2}} \Big) \Big(1 - i\Big) 
S_{n - 1} \nonumber \\
& & + \Big(\frac{1 + \gamma^{-2}}{4} \Big) \sum_{k = -5}^5
\Big(V_{\frac{|n - 2 k|}{2}} - V_{\frac{|n - 2 k| -2 }{2}} \Big)
~ A^{-2k}_{n - 2k} ~ S_{n - 2k}
\end{eqnarray}
The definitions of $k^{\pm}_n$ and $A^l_n$ are given in \cite{IsoCosmo}. 

Observe that we must have infinitely many $S_n$ to be non-zero for a
non-trivial solution and the number of non-trivial independent
solutions is reduced by one in both cases since the coefficient of
$S_0$ is zero (see \cite{DynIn}). We are interested in the
admissibility of a continuum approximation and in particular in
checking if the evolution is locally stable.

To explore these properties, we consider the large $n$ behavior of the 
coefficients, $S_n$. For definiteness we take $n$ to be positive and large 
which gets rid of the absolute values in the eigenvalues of the volume 
operator. Introduce the notation: 
\begin{equation}
a^2(n) := \frac{\gamma \lP^2 n}{6}
\end{equation}

For large $n$ one gets,
\begin{eqnarray}
V_{\frac{n + b}{2}} & \longrightarrow & a^3 \Big( 1 + \frac{3}{2}\frac{b +
1}{n} + \frac{3 b^2 + 6 b - 1}{8 n^2} + \cdots \Big) \\
k^{\pm}_n & \longrightarrow & 1 + o\Big(\frac{1}{n}\Big) 
\end{eqnarray}

For Bianchi-IX one needs $A^l_n$ defined in terms of $k^q_n :=
({\cal{K}}^q_{n+1} - {\cal{K}}^q_{n-1})/2$ for $q = \pm1, \pm3, \pm5$.
These become,
\begin{eqnarray}
{\cal{K}}^{\pm1}_{n + b} & \longrightarrow & \mp \Big(\frac{1 \mp
i}{36}\Big) \Big\{ \mp 9 n + \frac{9}{4} \mp 9b ~ +
~o\Big(\frac{1}{n}\Big) \Big\} \\
{\cal{K}}^{\pm3}_{n + b} & \longrightarrow & \pm \Big(\frac{5 \mp
i}{36}\Big) \Big\{ \mp \frac{27}{2} n + \frac{81}{8} \mp \frac{27}{2} b ~ +
~o\Big(\frac{1}{n}\Big) \Big\} \\
{\cal{K}}^{\pm5}_{n + b} & \longrightarrow & \mp \Big(\frac{1 \pm
i}{36}\Big) \Big\{ \mp \frac{45}{2} n + \frac{225}{8} \mp \frac{45}{2} b ~ +
~o\Big(\frac{1}{n}\Big) \Big\} \\
k^{\pm1}_{n + b} & \longrightarrow & + \Big( \frac{1 \mp i}{4} \Big) ~ + ~
o \Big( \frac{1}{n} \Big) \\
k^{\pm3}_{n + b} & \longrightarrow & - \Big( \frac{3 ( 5 \mp i )}{8}
\Big) ~ + ~ o \Big( \frac{1}{n} \Big) \\
k^{\pm5}_{n + b} & \longrightarrow & + \Big( \frac{5 ( 1 \pm i )}{8}
\Big) ~ + ~ o \Big( \frac{1}{n} \Big) 
\end{eqnarray}

With these, the difference equations become:
\begin{eqnarray}
0 ~ & = & ~ S_{n + 4} - 2 S_n + S_{n - 4} - \frac{1 + \gamma^{-2}}{4} (
S_{n + 8} - 2 S_n + S_{n - 8} ) ~~~~~~~\mbox{(Bianchi-I)} \\
0 ~ & = & ~ (1 - i)S_{n + 5} + (1 + i)S_{n - 5} + (1 + 5i)S_{n + 3} + (1
- 5i)S_{n - 3} - 2(1 + i)S_{n + 1} \nonumber \\
& & - 2(1 - i)S_{n -1} + \Big( \frac{1 + \gamma^{-2}}{2} \Big)\Big( 
\sum_{k = -5}^{5} A^{-2k}_{n - 2k} S_{n - 2k} \Big) ~~~~~~~~~~~~ 
\mbox{(Bianchi-IX)}
\end{eqnarray}

The coefficients $A^m_n$ satisfy the unitarity conditions $A^{-m}_n =
(A^m_n)^*$ and are obtained as:
\begin{eqnarray}
A^{10}_n & ~ = ~ & + \frac{25}{32}i  \nonumber \\
A^{8}_n & ~ = ~ & - \frac{45}{16} - \frac{15}{8}i  \nonumber \\
A^{6}_n & ~ = ~ & + 4 - \frac{45}{32}i  \nonumber \\
A^{4}_n & ~ = ~ & - \frac{3}{4} + \frac{7}{4}i  \nonumber \\
A^{2}_n & ~ = ~ & - 3 - \frac{59}{16}i  \nonumber \\
A^{0}_n & ~ = ~ & + \frac{73}{8}
\end{eqnarray}

Note that in the large-$n$ limit, the equation has reduced to a difference
equation with constant coefficients and our first necessary condition
for admissibility of continuum approximation is satisfied. The condition
of Eq.\ (\ref{CoeffCond}) however is satisfied by the Bianchi-I case but
{\em not} by the Bianchi-IX case. Therefore we will not get the
Wheeler--DeWitt equation in the leading approximation for the latter.

A simple check on local stability is to obtain the characteristic
roots to the {\em leading order in $\gamma$}. This picks out the polynomial
obtained from the coefficients in the term multiplying $\gamma^{-2}$.

In both cases of Bianchi types, the corresponding characteristic  polynomial 
is a polynomial in $z^2$. Hence both $\pm z$ are roots. The coefficients of 
the polynomial also come in complex conjugate pairs such that if $z$ is a root
then so is $1/z^*$. Furthermore, these polynomials are perfect squares, so 
each root is a {\it double} root. Corresponding to each such root, there will 
be solutions which will behave as $z^n(a n + b)$.

For Bianchi-I, in the continuum approximation, the roots of the
characteristic polynomial are the $8^{th}$ roots of unity each being a
double root.  Clearly all those roots have absolute values equal to 1
and exactly one root equals 1. Our condition of local stability is
thus satisfied (see also \cite{SemiClass} for the continuum limit). 

For Bianchi-IX on the other hand, the roots are explicitly obtained as:
\begin{eqnarray}
\lambda_0 & = & 0.24077449441476 - i ~ 0.47063359530671 ~~~,~~~ |\lambda_0| =
0.52864765032298 \nonumber \\
\lambda_1 & = & 0.29768070243356 + i ~ 0.95466549083889 ~~~,~~~ |\lambda_1| = 
1.0 \nonumber \\
\lambda_2 & = & 0.8 - i ~ 0.6
~~~~~~~~~~~~~~~~~~~~~~~~~~~~~~~~~~~~~~~~~~,~~~ |\lambda_2| = 
1.0 \nonumber \\
\lambda_3 & = & 0.86154480315169 - i ~ 1.68403189553218 ~~~,~~~ |\lambda_3| =
1.89161911414729  \nonumber \\
\lambda_4 & = & -1.0  ~~~~~~~~~~~~~~~~~~~~~~~~~~~~~~~~~~~~~~~~~~~~~~~~~,~~~
|\lambda_4| = 1.0 
\end{eqnarray}

The actual 10 double roots are $z^{\pm}_i = \pm \sqrt{\lambda_i}$.
Observe that 1 is not a root of the characteristic polynomial. Hence
the approximating differential equation equation will also have lower
derivative terms. Moreover we do have roots whose absolute values are
not equal to one, thus violating local stability.

Thus, the original quantization in \cite{IsoCosmo} for the closed
isotropic model is not admissible by our criteria. In \cite{Closed},
a different quantization of the closed model has been derived with the
methods of \cite{HomSpin}, providing an admissible quantization for
this model.

%%%%%%%%%%%%%%%%%%%%%%%%%%%%%%%%%%%%%%%%%%%%%%%%%%%%%%%%%%%%%%%%%%%%%%%

\end{document}